\documentclass{article}
\usepackage{graphicx}
\usepackage{amsmath}

\newcommand{\middlefig}{.5\textwidth}

\begin{document}
 
\baselineskip 18pt

\begin{center}
{\Large {\bf  Modulational Instability in Bose-Einstein Condensates under Feshbach Resonance Management}}
 \vskip5mm Z. Rapti$^{1}$,
G. Theocharis$^{2}$, P.G. Kevrekidis$^{1}$, D.J. Frantzeskakis$^{2,}$\footnote{Email address: dfrantz@cc.uoa.gr}, and B.A. Malomed$^3$ \vskip3mm \mbox{}%
$^{1}$ Department of Mathematics and Statistics, University of Massachusetts,
Amherst MA 01003-4515, USA 

\mbox{}$^{2}$ Department of Physics, University of Athens, 
Panepistimiopolis, Zografos, Athens 15784, Greece

\mbox{}$^{3}$ Department of Interdisciplinary Studies, Faculty of Engineering, 
Tel Aviv University, Tel Aviv 69978, Israel

\bigskip

\begin{abstract}
We investigate the modulational instability of nonlinear Schr{\"o}dinger
equations with periodic variation of their coefficients. In particular, we
focus on the case of the recently proposed, experimentally realizable
protocol of Feshbach Resonance Management for Bose-Einstein condensates. We
derive the corresponding linear stability equation analytically and we show
that it can be reduced to a Kronig-Penney model, which allows the
determination of the windows of instability. The results are tested
numerically in the absence, as well as in the presence of the magnetic
trapping potential.

\bigskip
\noindent PACS numbers: 03.75.-b, 03.75.Kk, 03.75.Lm 
\end{abstract}
\end{center}

\newpage

\section*{1. Introduction}

The experimental realization and intense theoretical studies of
Bose-Einstein condensates (BECs) \cite{review} have led to an explosion of
interest in the field of atomic matter waves and nonlinear excitations in
them. In particular, one-dimensional (1D) dark \cite{dark} and bright \cite
{bright} solitons have been created in experiments, in BECs with repulsive
and attractive interactions, which correspond to the positive and negative
atom-atom scattering length (SL), respectively. Additionally,
multi-dimensional solitons \cite{BaizOstr}, Faraday waves \cite{stal}, ring
dark solitons and vortex necklaces \cite{theo}, have been theoretically
predicted. The BEC solitons are of fundamental interest, not only
conceptually, but also practically: one can envision robust solitary
structures being coherently manipulated in matter-wave chips, similar to how
light is controlled in the existing optical ones \cite{Folman}.

To generate BEC solitons in one spatial dimension (in the presence of the
magnetic trap \cite{us,konotop}, or in the quasi-discrete setting created by
an optical-lattice potential \cite{abdul2}), as well to avoid their collapse
in two dimensions \cite{abdul1}, an experimentally realizable protocol has
been recently proposed, in the form of the so-called Feshbach Resonance
Management (FRM). It is based on adjusting the effective SL (including a
possibility to change its sign) by a resonantly tuned ac magnetic field
through the Feshbach resonance \cite{inouye}.

The FRM scheme can be modelled (in the mean-field approximation) in the
framework of the Gross-Pitaevskii (GP) equation \cite{review}, with the
coefficient in front of the nonlinear term being a periodic function of
time. In Ref. \cite{us}, the periodic function has been taken to be a
piece-wise constant one, periodically jumping between positive and negative
values. The same model may also be realized in terms of nonlinear optics,
where it applies to a medium composed of alternating layers with opposite
signs of the Kerr nonlinearity \cite{Isaac}. FRM resembles the \textit{
dispersion-management} (DM) scheme, well-known in fiber optics (see, e.g., 
\cite{Progress} and \cite{turits} for review), which assumes that nonlinear
fibers with opposite signs of the group-velocity dispersion periodically
alternate, to form a system supporting robust breathing solitons. Similarly
to this, very robust and stable matter-wave breathing solitons can be
generated by means of the FRM in BECs \cite{us}, which lead to structures
(such as, e.g., the dark solitons) even more robust than their optical
counterparts \cite{pare}.

The purpose of this paper is to study how solitary waves can arise in the
FRM setting and, in particular, to investigate the modulational instability
(MI) in the corresponding nonlinear Schr{\"{o}}dinger (NLS) equations with
periodically varying coefficients. Such a study is relevant, as MI is one a
fundamental mechanism that leads to the formation of localized solitary-wave
structures in a variety of settings, ranging from fluid dynamics (where it
is usually referred to as the Benjamin-Feir instability) \cite{benjamin67}
to nonlinear optics \cite{ostrovskii69} and plasma physics \cite{taniuti68},
and, most recently, to BECs in the presence of an optical lattice \cite
{wu01} or magnetic-trap potentials \cite{ourpra}. In the
present work, the study of the MI leads to a general linear stability
equation which includes, as special cases, equations previously examined in
the context of DM. We then proceed to analyze MI conditions in the case of
the piecewise-constant FRM scheme \cite{us}, and find that the resulting
stability equation resembles the well-known Kronig-Penney model of solid-state
physics \cite{Kittel}. Thus, we derive analytical criteria for the MI.

The paper is organized as follows. The derivation of the stability equation
and its analytical treatment are presented in Section 2. In Section 3, we
turn to numerical experiments both in the absence of the magnetic trap (the
case for which we have analytical results) and in the presence of it, as the
magnetic trap is a necessary ingredient of the experiment. In Section 4 we
summarize our findings and present our conclusions.

\section*{2. Stability analysis}

We consider a general GP/NLS equation, with time-dependent coefficients in
front of the dispersive and the nonlinear terms, $D(t)$ and $a(t)$,
respectively. The equation also includes a term corresponding to the
external magnetic (parabolic) potential trap:
\begin{equation}
iu_{t}=-D(t)u_{xx}+a(t)|u|^{2}u+(1/2)\Omega ^{2}x^{2}u.  \label{fmeq1}
\end{equation}
In the case of BECs ($D=\mathrm{const}\equiv 1/2$), $u(x,t)$ is the
macroscopic wave function, $x$ and $t$ are measured in the
harmonic-oscillator units, and $\Omega $ is the strength of the magnetic
trap \cite{review}. If the FRM is applied \cite{us}, the nonlinear
coefficient $a(t)$ periodically alternates between $a_{1}$ (for $0<t\leq
\tau $) and $-a_{2}$ (for $\tau <t\leq T$), where it is assumed that 
$a_{1},a_{2}>0$. As Eq. (\ref{fmeq1}) is spatially inhomogeneous due to the
presence of the parabolic trap, analytical investigation of the linear
stability of the uniform states is only possible for $\Omega =0$. Hence, we
will examine this case, which will be complemented by numerical results for
both $\Omega =0$ and $\Omega \neq 0$.

The plane-wave solution to Eq. (\ref{fmeq1}) is 
\begin{equation*}
u_{0}=A_{0}\exp \left[ i(-q^{2}\int_{0}^{t}D(s)ds-A_{0}^{2}
\int_{0}^{t}a(s)ds+qx)\right] .
\end{equation*}
Notice that the solution's amplitude can be rescaled to $A_{0}\equiv 1$.
Then, a solution including an infinitesimal perturbation is sought as 
\begin{equation*}
u=u_{0}\left[ 1+\epsilon w(t)\cos (kx)\right] ,
\end{equation*}
where $\epsilon $ and $k$ are the amplitude and wavenumber of the
perturbation, which leads to the following linearized equations for $w\equiv
w_{r}+iw_{i}$: 
\begin{equation*}
\dot{w}_{r}=k^{2}D(t)w_{i},~\dot{w}_{i}=-\left[ k^{2}D(t)+2a(t)\right] w_{r},
\end{equation*}
which can be transformed into a single equation, 
\begin{equation}
\ddot{w}_{r}=\dot{D}D^{-1}\dot{w}_{r}-k^{2}D(t)\left[ k^{2}D(t)+2a(t)\right]
w_{r},  \label{fmeq2}
\end{equation}
the overdot standing for $d/dt$.

There are several special cases of this equation that were previously
studied. In the case of DM (i.e., for $D=D(t)$ and $a(t)\equiv \mathrm{const}
$), the MI analysis was performed in \cite{doran}. On the other hand, in the
FRM context, for $D\equiv 1/2$ and time-periodic $a(t)$, Eq. (\ref{fmeq2})
is a Hill equation that was considered in \cite{us}, while the more
specialized case of $a(t)=1+2\alpha \cos (\omega t)$ leads to the Mathieu
equation that was dealt with, in this context, in \cite{stal}.

Motivated by the FRM scheme proposed in \cite{us}, we will now explore the
special case of a piecewise-constant time-dependent SL (alternating between 
$a_{1}$ and $-a_{2}$ as discussed above). In this case, we define 
$s_{1}^{2}=k^{2}(k^{2}/4+a_{1})$ and $s_{2}^{2}=k^{2}(a_{2}-k^{2}/4)$, which
assumes $k^{2}<4a_{2}$. Searching for a solution in accordance with Bloch's
theorem \cite{Kittel}, we find, within one period, 
\begin{eqnarray}
w_{r} &=&Ae^{i(-\omega +s_{1})t}+Be^{i(-\omega -s_{1})t},~~~\mathrm{for}
~0<t\leq \tau ,  \label{hill1} \\
w_{r} &=&Ce^{(-i\omega +s_{2})t}+De^{(-i\omega -s_{2})t},~~\mathrm{for}~\tau
\leq t\leq T.  \label{hill2}
\end{eqnarray}
Then one should request the continuity of the solution and its derivative at 
$t=0$ and $t=\tau $, obtaining 4 conditions on $A,B,C,D$, namely, 
\begin{eqnarray}
e^{is_{1}\tau }A+e^{-is_{1}\tau }B-e^{s_{2}\tau }C-e^{-s_{2}\tau }D &=&0, 
\notag  \label{hill3} \\
i(-\omega +s_{1})e^{is_{1}\tau }A+i(-\omega -s_{1})e^{-is_{1}\tau
}B-(-i\omega +s_{2})e^{s_{2}\tau }C-(-i\omega -s_{2})e^{-s_{2}\tau }D &=&0, 
\notag  \label{hill4} \\
A+B-e^{(-i\omega +s_{2})T}C-e^{(-i\omega -s_{2})T}D &=&0,  \notag
\label{hill5} \\
i(-\omega +s_{1})A+i(-\omega -s_{1})B-(-i\omega +s_{2})e^{(-i\omega
+s_{2})T}C-(-i\omega -s_{2})e^{(-i\omega -s_{2})T}D &=&0.  \notag
\label{hill6}
\end{eqnarray}

The resulting system of homogeneous linear equations for $(A,B,C,D)$ has
non-trivial solutions only if its determinant vanishes. This condition leads
to the following equation for the eigenfrequency (Floquet multiplier) 
$\omega $: 
\begin{equation}
\cos (\omega T)=-\frac{s_{1}^{2}-s_{2}^{2}}{2s_{1}s_{2}}\sin (s_{1}\tau
)\sinh [s_{2}(T-\tau )]+\cos (s_{1}\tau )\cosh [s_{2}(T-\tau )]\equiv F(k).
\label{hill7}
\end{equation}
If the above condition $k^{2}<4a_{2}$ does not hold, we redefine $\tilde{s}
_{2}=\sqrt{k^{2}(k^{2}/4-a_{2})}$, and obtain, instead of Eq. (\ref{hill7}), 
\begin{equation}
\cos (\omega T)=-\frac{s_{1}^{2}+\tilde{s}_{2}^{2}}{2s_{1}\tilde{s}_{2}}\sin
(s_{1}\tau )\sin [\tilde{s}_{2}(T-\tau )]+\cos (s_{1}\tau )\cos [\tilde{s}
_{2}(T-\tau )]\equiv \widetilde{F}(k).  \label{hill8}
\end{equation}

By examining the function $\left\vert F(k)\right\vert $ or $\left\vert 
\widetilde{F}(k)\right\vert $, defined in Eqs. (\ref{hill7}) and 
(\ref{hill8}), and comparing it to $1$, we can find whether there is a real
eigenfrequency for a given perturbation wavenumber $k$, or it belongs to a
\textquotedblleft forbidden zone\textquotedblright , which implies the MI.

\begin{figure}[!hbp]
\begin{center}
    \includegraphics[width=\middlefig]{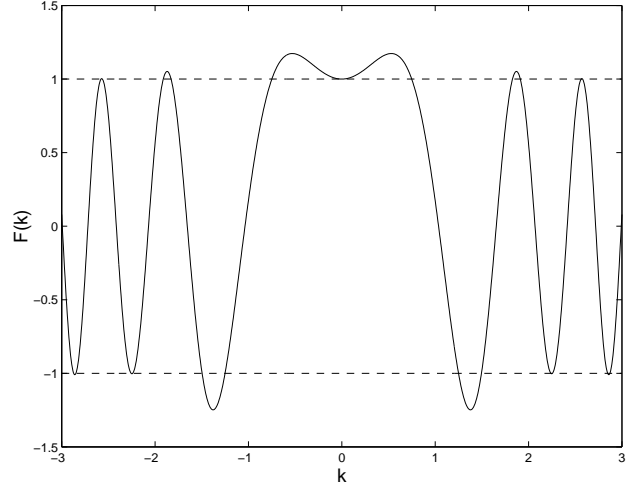}
\bigskip 
\end{center}
\caption{The plot of the function $F(k)$ defined by Eqs. (\ref{hill7}) and 
(\ref{hill8}) for $a_1=a_2=0.3$ and $\protect\tau= T/4= 1$. When the function
satisfies $|F(k)| \leq 1$, the perturbation of wavenumber $k$ is
modulationally stable, while for $|F(k)|>1$, it is modulationally unstable.}
\label{fig1}
\end{figure}

A typical example of results produced by this analysis is shown in Fig. \ref
{fig1}, for $a_{1}=a_{2}=0.3$ and $\tau =T/4=1$. The first several
instability windows, which are determined by the shape of the curve $F(k)$
for this case are given in Table I. Notice that in the table, only positive
values of unstable wavenumbers are given, the negative ones obeying the 
$k\rightarrow -k$ symmetry. In general, we have found that the number and widths of the instability windows increase as long as the mean value $\bar{a}\equiv \lbrack a_{1}\tau
-a_{2}(T-\tau )]/T$ of the SL gets large and negative, i.e., when the BEC is
``very attractive'' on the average. 

\section*{3. Numerical Results}

We now examine the validity of the above theoretical predictions through
full numerical simulations of Eq. (\ref{fmeq1}). This also allows us to
investigate the role of the magnetic trap that was not taken into regard in
the above analysis.

As a typical example (similar results have been obtained in other cases), we
consider modulationally stable and unstable cases for the example presented
above (i.e., $a_{1}=a_{2}=0.3$ and $\tau =T/4=1$.). In particular, we
examine the case of $k=0.5$ which, according to Fig. \ref{fig1} and Table I,
should be unstable, and the case of $k=1$, that is expected to be stable. It
can be clearly observed that the corresponding configurations evolve in
accordance with the theoretical prediction, indeed leading to the MI and, as
a result, generation of a lattice of solitary waves, for $k=0.5$. Instead,
the modulational perturbations remain small for $k=1$. The agreement between
the analytical predictions and direct simulations has been found to be
generic.

We now proceed to the case when a weak magnetic trap is included, taking
(for example) $\Omega ^{2}=0.00025$. Then we observe that \textit{both} the
configurations which were linearly stable and unstable in the absence of the
trap \textit{eventually} develop the MI, which is in consonance also with
the findings of \cite{ourpra}. Here, we used the initial condition $u=u_{
\mathrm{TF}}[1+0.02\cos (kx)]$, where $u_{\mathrm{TF}}=\sqrt{\max (0,\mu
-V(x))}$ is the wave function in the Thomas-Fermi approximation 
\cite{review}. The reason why the MI develops in both cases 
is that the magnetic trap
mixes stable and unstable wavenumbers in the Fourier space. As a result,
even though the initial configuration contained only stable wavenumbers,
unstable ones are eventually generated, giving rise the MI. However, as can
be observed in Fig. \ref{fig3}, the MI occurs \textit{faster} in the case that
would be modulationally unstable in the absence of the trap, as the MI
growth rate is larger in this case than in the case when the uniform state
was stable in the absence of the trap.

We also examined what happens with the increase of the strength of the
magnetic trap, which makes it \textquotedblleft tighter\textquotedblright.
In this case, the trapping potential is much more efficient in mixing the
wavenumbers, therefore there remains little difference between the evolution
of configurations that were modulationally unstable and stable without the
trap. An example is shown in Fig. \ref{fig4} for $\Omega ^{2}=0.0015$ ($6$
times as large as in the case shown in Fig. \ref{fig3}). It can be seen that
the development of the MI for $k=0.5$ and $k=1$ is practically identical.

\section*{4. Conclusions}

In this paper, we have studied the modulational stability (MI) in the GP/NLS
equations with periodically varying coefficients. We highlighted how the
results may be applied to problems stemming from both nonlinear optics and
Bose-Einstein condensation. Findings reported in earlier works in the
subject have been generalized. The equation obtained from the analysis, Eq. 
(\ref{fmeq2}), is of interest in its own right. Motivated by the recently
introduced concept of a periodically modulated (via a Feshbach resonance)
scattering length as a means of generating robust breathing matter-wave
solitons \cite{us}, we have focused attention on the case of alternating
positive and negative scattering length.

For the case of the FRM, we have demonstrated that the above-mentioned equation
governing the MI development is an equation of the Hill type. More
specifically, it is an analog of the Kronig-Penney model known in solid
state physics. Then, developing analysis similar to the eigenvalue
calculation in the Kronig-Penney model, we have obtained analytical
conditions for the MI of such FRM schemes. The analytical criteria have been
shown to be in agreement with full numerical simulations. Although the
analysis was carried out in the absence of the magnetic trap, the role of
the trap was investigated numerically. For weak trap potentials, it was
found that the MI will always set in (due to the mixing of wavenumbers
imposed by the spatial potential); however, its growth rate bears
``memory'' of the presence/absence of the
MI in the corresponding situation without the trap. On the other hand, if
the trap is strong, the memory is effectively wiped out.


\bigskip\noindent {\large \textbf{Acknowledgements}}

This work was partially supported by a UMass FRG, NSF-DMS-0204585 and the
Eppley Foundation for Research (PGK), as well as by the Special Research
Account of the University of Athens under grant No. 70/4/5844 (GT, DJF).

\newpage

\vspace{5mm} 
\begin{tabular}{|l|l|}
\hline
\centering Window & k \\ \hline
1 & [0,0.7489) \\ \hline
2 & (1.2502,1.4957) \\ \hline
3 & (1.8283,1.9115) \\ \hline
4 & (2.2396,2.2562) \\ \hline
5 & (2.5638,2.5784) \\ \hline
\end{tabular}
\newline
\newline
Table I: The first instability bands of the perturbation wavenumber $k$ for
the case $a_1=a_2=0.3$ and $\tau=T/4=1$. 



\begin{figure}[!hbp]
\begin{center}
\begin{tabular}{cc}
    (a) & (b) \\
    \includegraphics[width=\middlefig]{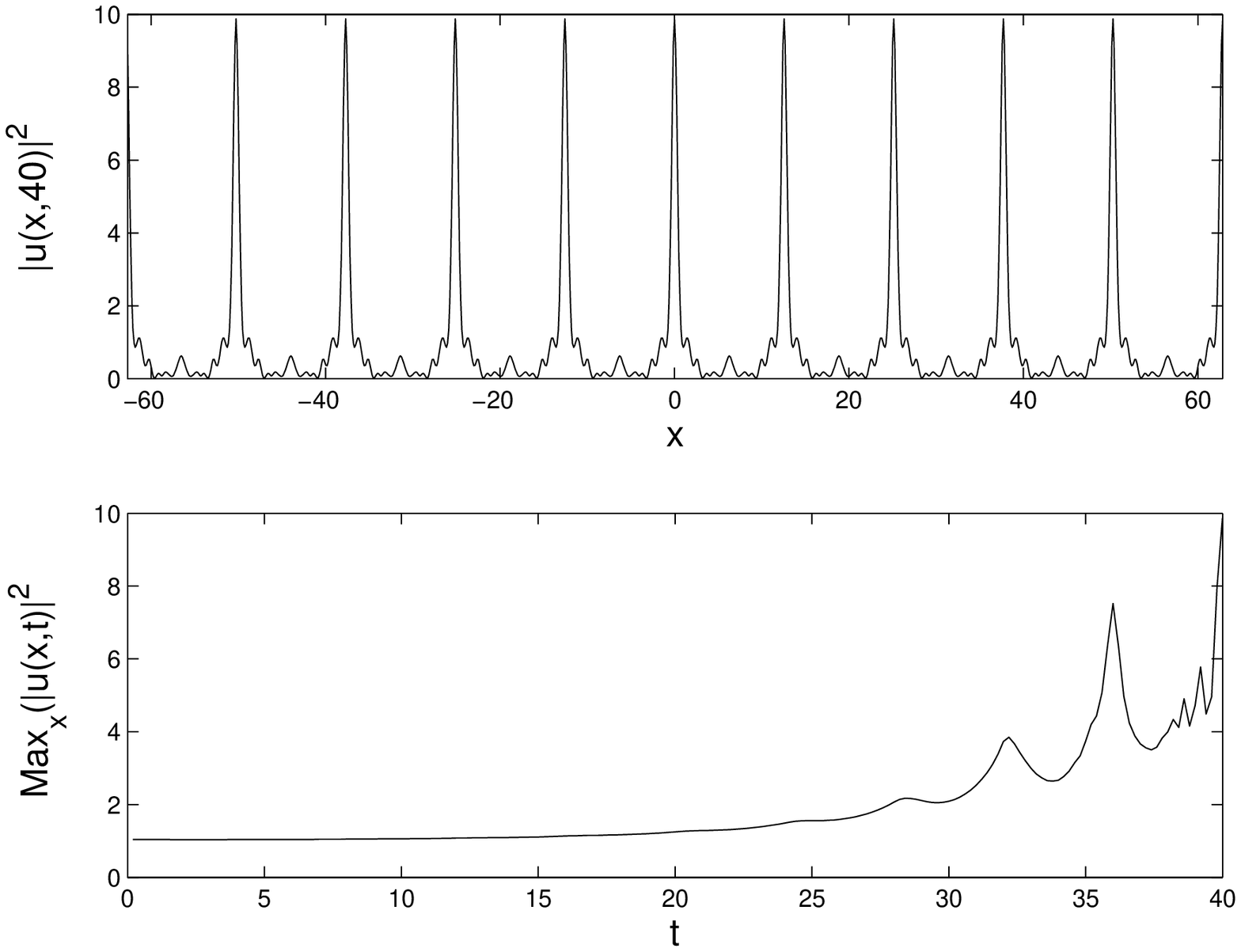}
    \includegraphics[width=\middlefig]{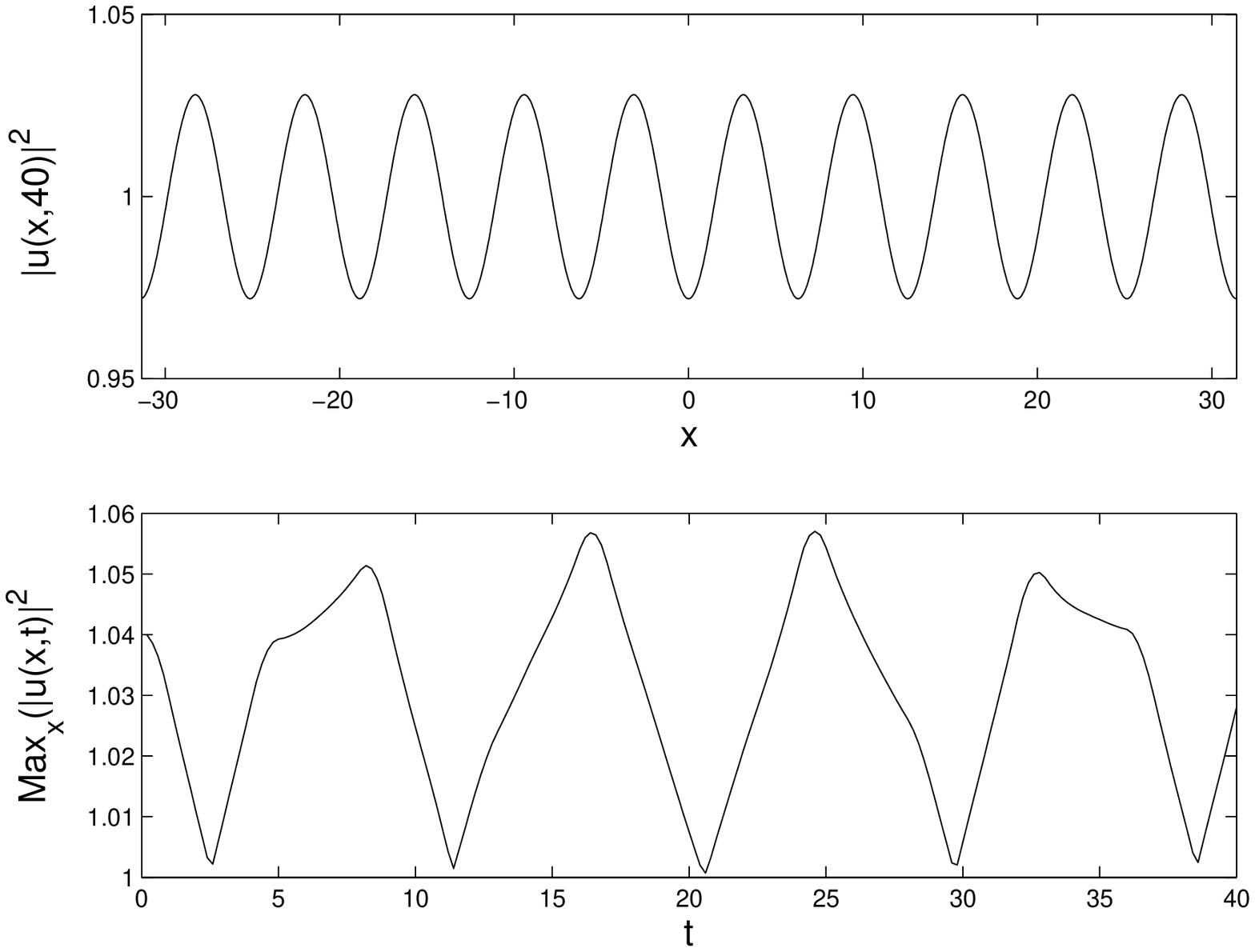}
\end{tabular}
\bigskip 
\end{center}
\caption{For the case $a_1=a_2=0.3$ and $\protect\tau= T/4= 1$, $\Omega=0$,
the evolution of modulationally unstable ($k=0.5$, left panel) and stable 
($k=1$, right panel) perturbations of the form $0.02 \cos(kx)$ added to $u=1$,
is examined. The top and bottom panels show, respectively, the configuration
at $t=40$ and the time evolution of the amplitude of the spatial profile.}
\label{fig2}
\end{figure}

\begin{figure}[!hbp]
\begin{center}
\begin{tabular}{cc}
    (a) & (b) \\
    \includegraphics[width=\middlefig]{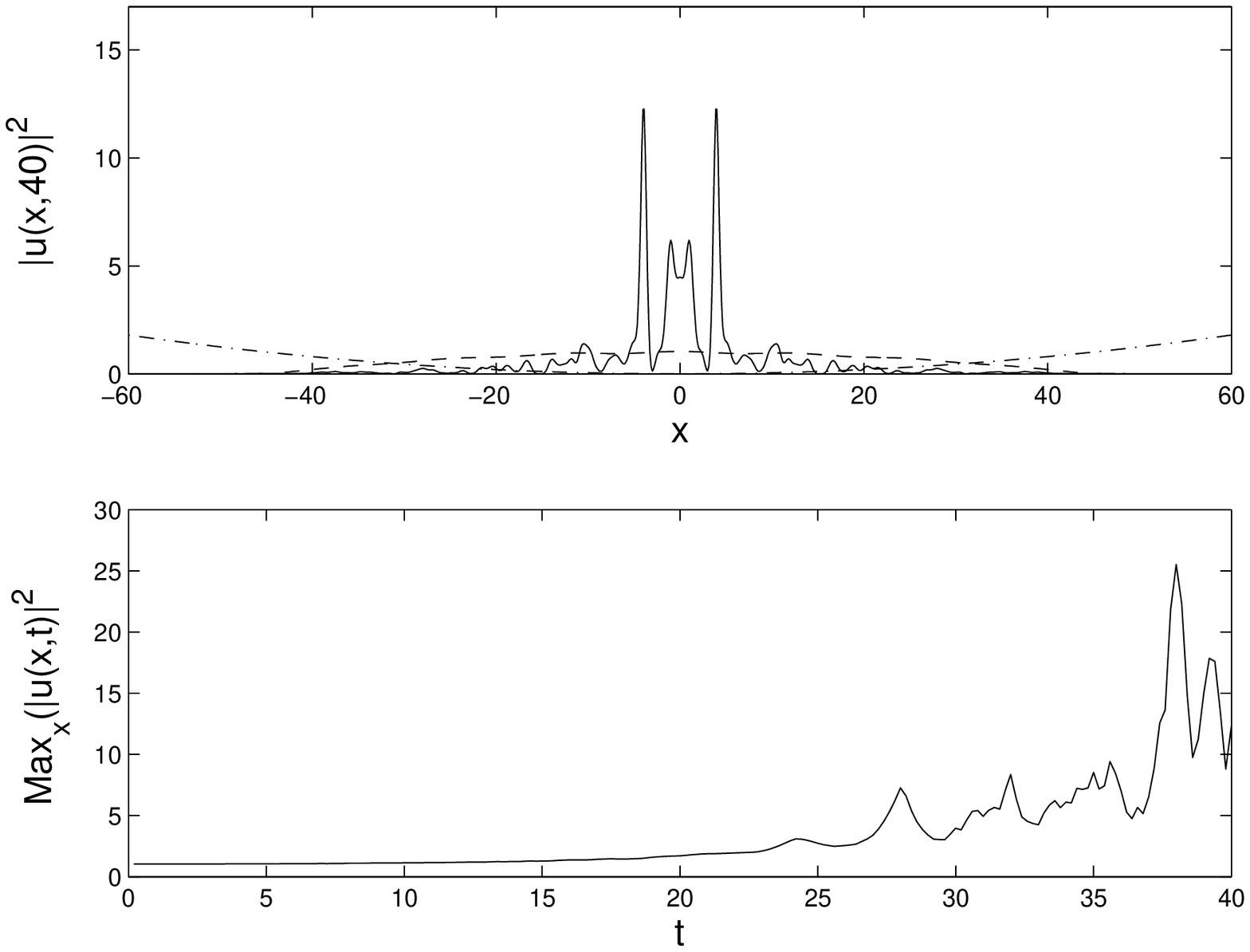}
    \includegraphics[width=\middlefig]{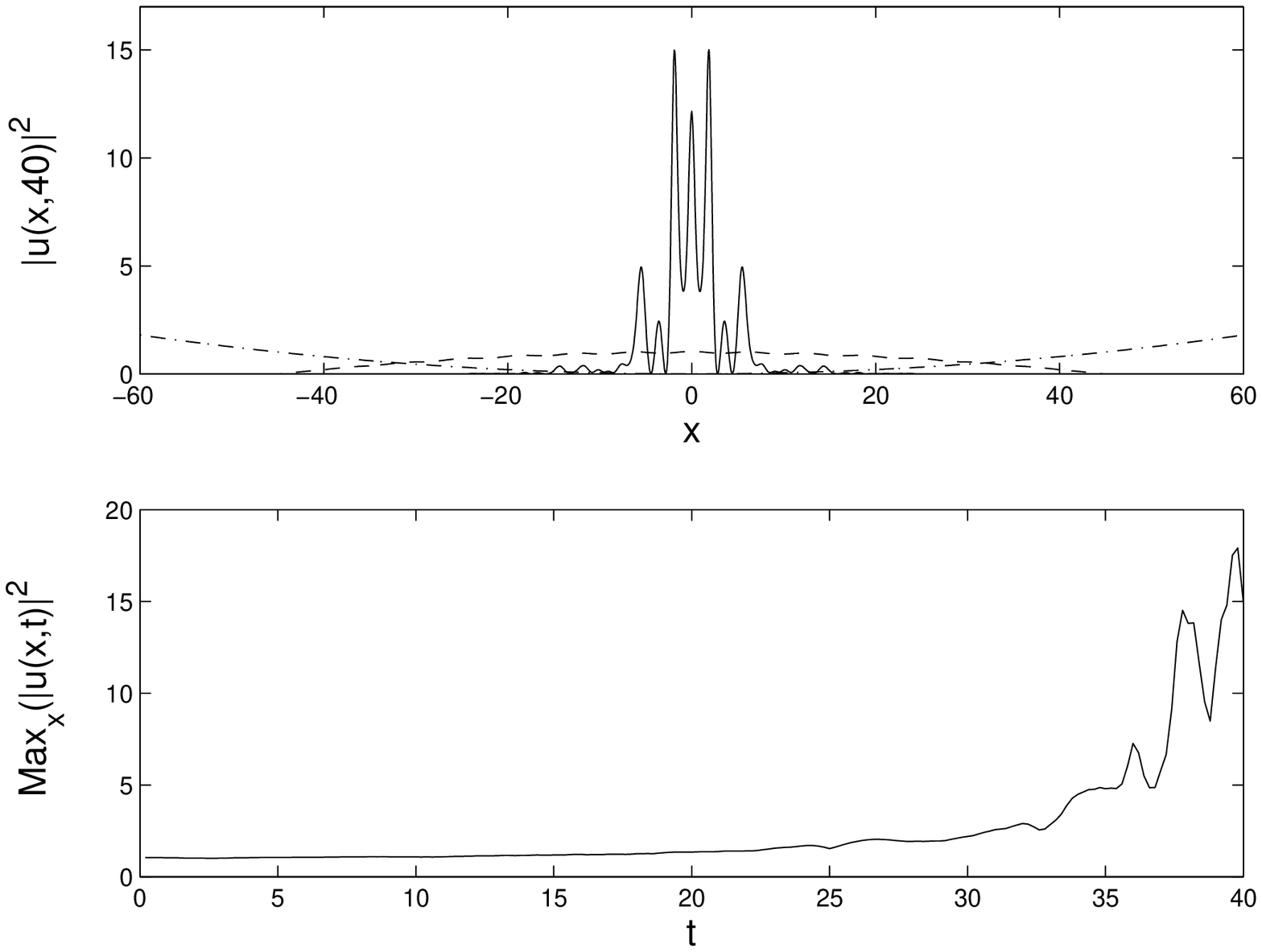}
\end{tabular}
\end{center}
\caption{The same as in the previous figure, but now with 
$\Omega^2=0.00025$, the initial configuration being multiplied 
by the Thomas-Fermi wave
function $\protect\sqrt{\max(0,\protect\mu - V(x))}$ with $\protect\mu=1$.
Notice that the modulational instability develops both for $k=0.5$ (left)
and for $k=1$ (right).}
\label{fig3}
\end{figure}

\begin{figure}[!hbp]
\begin{center}
\begin{tabular}{cc}
    (a) & (b) \\
    \includegraphics[width=\middlefig]{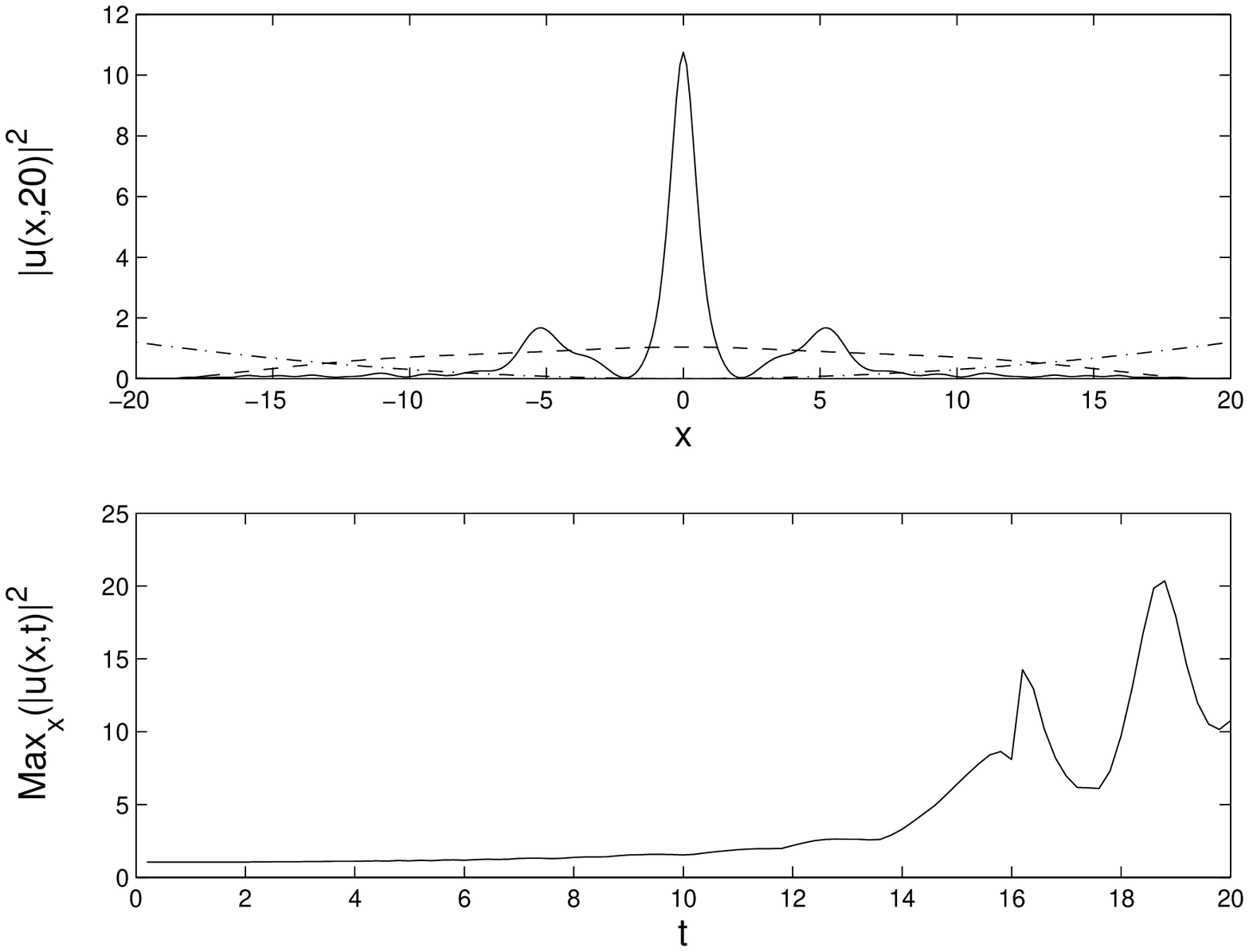}
    \includegraphics[width=\middlefig]{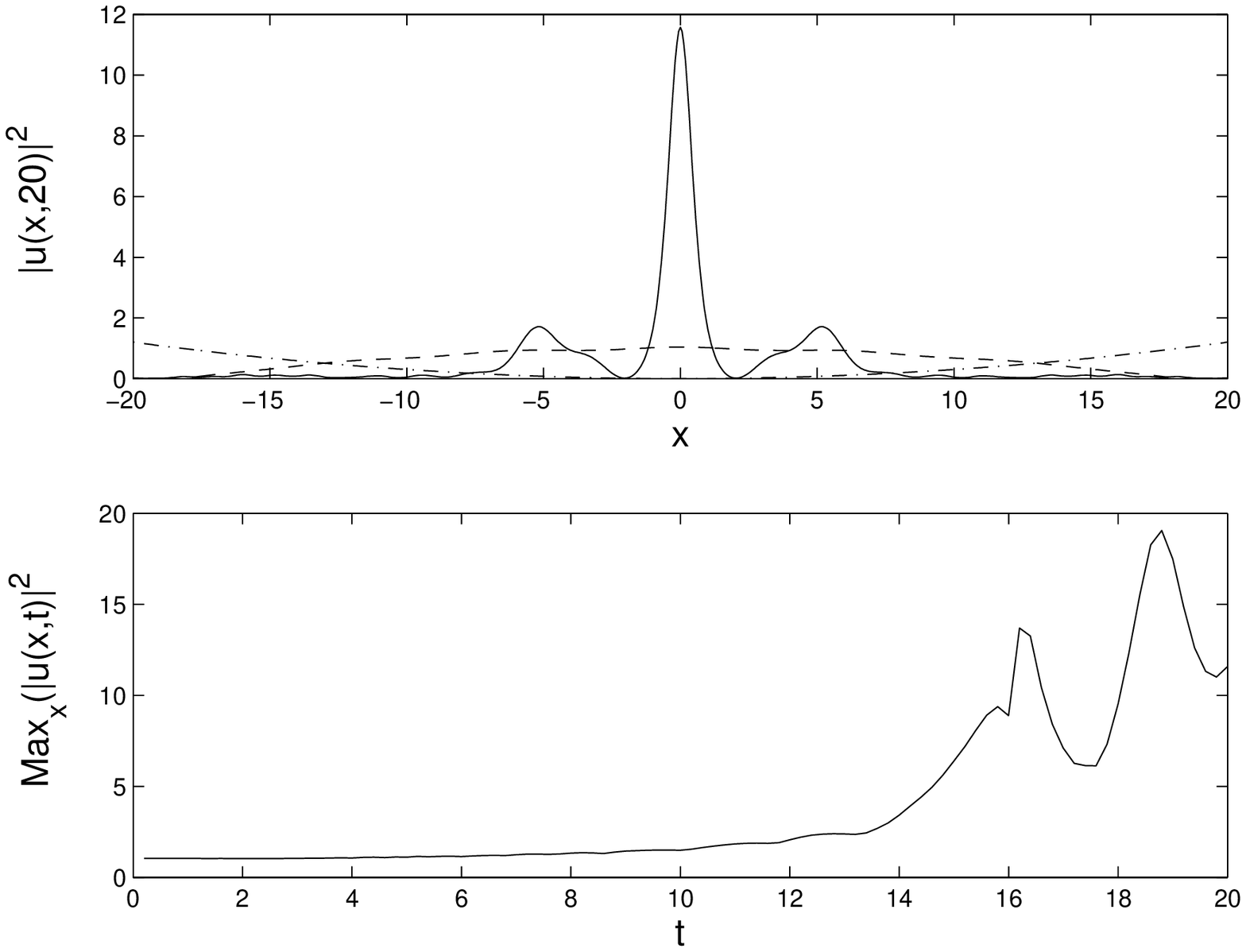}
\end{tabular}
\end{center}
\caption{The same as in the previous figure, but with $\Omega^2=0.0015$. The
only remaining visible difference between the cases of $k=0.5$ (left) and 
$k=1 $ (right) is a slightly larger growth rate in the former case.}
\label{fig4}
\end{figure}


\begin{thebibliography}{99}
\bibitem{review} F. Dalfovo \textit{et al.}, Rev. Mod. Phys. \textbf{71},
463 (1999).

\bibitem{dark} S. Burger \textit{et al.}, Phys. Rev. Lett. \textbf{83},
5198(1999); J. Denschlag \textit{et al.}, Science \textbf{287}, 97 (2000);
B. P. Anderson \textit{et al.}, Phys. Rev. Lett. \textbf{86}, 2926 (2001).

\bibitem{bright} K.E. Strecker \emph{et al.}, Nature \textbf{417}, 150
(2002); L. Khaykovich \emph{et al.}, Science \textbf{296}, 1290 (2002).

\bibitem{BaizOstr} B.B.\ Baizakov, V.V.\ Konotop, and M.\ Salerno, J.\
Phys.\ B \textbf{35}, 5105 (2002); E.A.\ Ostrovskaya and Y.S.\ Kivshar,
Phys.\ Rev.\ Lett.\ \textbf{90}, 160407 (2003).


\bibitem{stal} K. Staliunas, S. Longhi and G. J. de Valc\'{a}rcel, \newblock
Phys. Rev. Lett. \textbf{89}, 210406 (2002).

\bibitem{theo} G. Theocharis \textit{et al.}, Phys. Rev. Lett. \textbf{90},
120403 (2003).

\bibitem{Folman} R.\ Folman and J.\ Schmiedmayer, Nature \textbf{413}, 466
(2001); R.\ Folman \textit{et al.}, Adv.\ Atom.\ Mol.\ Opt.\ Phys.\ \textbf{
48}, 263 (2002).

\bibitem{us} P.G. Kevrekidis \textit{et al.}, Phys. Rev. Lett. \textbf{90},
230401 (2003).

\bibitem{konotop}  F.Kh. Abdullaev \textit{et al.}, Phys. Rev. Lett. 
\textbf{90}, 230402 (2003).

\bibitem{abdul2}  F.Kh. Abdullaev \textit{et al.}, cond-mat/0306281.

\bibitem{abdul1}  F. Kh. Abdullaev \textit{et al.}, Phys. Rev. A 
\textbf{67}, 013605 (2003); H. Saito and M. Ueda, Phys. Rev. Lett. 
\textbf{90}, 040403 (2003).

\bibitem{inouye}  S. Inouye \textit{et al.}, Nature \textbf{392}, 151
(1998); E.A. Donley \textit{et al.}, Nature \textbf{412}, 295 (2001).

\bibitem{Isaac}  I. Towers and B.A. Malomed, J. Opt. Soc. Am. B \textbf{19},
537 (2002).


\bibitem{Progress}  B.A.\ Malomed, Progress in Optics \textbf{43}, 71 (2002).

\bibitem{turits}  S.K. Turitsyn \textit{et al}, C.R. Physique \textbf{4},
145 (2003).


\bibitem{pare}  C. Par{\'{e}} and P.-A. B{\'{e}}langer, \newblock Opt.
Commun. \textbf{168}, 103 (1999).

\bibitem{benjamin67}  T.B. Benjamin and J.E. Feir, J. Fluid. Mech. 
\textbf{27}, 417 (1967).

\bibitem{ostrovskii69}  L.A. Ostrovskii, Sov. Phys. JETP \textbf{24}, 797
(1969).

\bibitem{taniuti68}  T. Taniuti and H. Washimi, Phys. Rev. Lett. 
\textbf{21}, 209 (1968); A. Hasegawa, Phys. Rev. Lett. \textbf{24}, 1165 (1970).

\bibitem{wu01}  B. Wu and Q. Niu, Phys. Rev. A \textbf{64}, 061603(R) (2001); 
V.~V.~Konotop, and M.~Salerno, Phys. Rev. A \textbf{65}, 021602 (2002); 
A. Smerzi \textit{et al} Phys. Rev. Lett., \textbf{89}, 170402 (2002).


\bibitem{ourpra}  G. Theocharis \textit{et al.}, Phys. Rev. A 67, 063610
(2003); L.D. Carr and J. Brand, cond-mat/0303257.

\bibitem{Kittel}  C. Kittel, \textit{Introduction to Solid State Physics}
(John Wiley and Sons: New York, 1986).



\bibitem{doran}  N.J. Smith and N.J. Doran, Opt. Lett. \textbf{21}, 570
(1996); F.Kh. Abdullaev \textit{et al}., Phys. Lett. A \textbf{220}, 213
(1996).





\end{thebibliography}
\end{document}